# A three-layer preon star model from exact piecewise-continuous solutions of Einstein's equations

Z. Pazameta

*Physical Sciences Department, Eastern Connecticut State University, Willimantic, CT 06226*

**Abstract**

A metric of the form $g_{tt} = f(r) = -1/g_{rr}$ is employed to model a hybrid astrophysical compact object consisting of a preon gas core, a mantle of electrically charged hot quark-gluon plasma, and an outer envelope of charged hadronic matter which is matched to an exterior Reissner-Nordstrøm vacuum. The piecewise-continuous metric and the pressure and density functions consist of polynomials that are everywhere well-behaved. Boundary conditions at each interface yield estimates for physical parameters applicable to each layer, and to the star as a whole.

# Introduction

In recent years, theoretical and experimental developments have opened up new avenues for pursuing the study of exotic astrophysical compact objects. One of these has been the formulation of the preon star concept by Hansson and Sandin [1] and the subsequent studies exploring this object's possible internal structure, observational properties and formation scenarios [2, 3, 4]. Another has come from insights into the physics of quark-gluon plasma (QGP) obtained via the Relativistic Heavy Ion Collider (RHIC) [5]. It occurred to us that combining these developments, within the framework of a physically plausible formation scenario, could yield a model for a compact object composed of several layers of material.

How could such an object form? The estimated masses for preon stars—including our hybrid model—are some three orders of magnitude too small for them to have formed by conventional post-stellar gravitational collapse. This means that, during the formation process, most of a progenitor star's mass must be ejected into space while a low-mass central region undergoes implosion to form the ultra-compact object. The theoretically favored means of producing a preon star, however, is through density fluctuations in the very early Universe; the chaotic nature of such fluctuations would prevent any stratification in the object's interior, so it would form as a homogenous sphere of pure preonic matter. Moreover, any such objects that survive today must have lost only tiny amounts of energy through interaction with their ambient over a long period of time—but, as discussed in Section 5, such energy loss could eventually lead to the formation of the stratified structure described below.

Consider an imploding post-stellar object whose internal density significantly exceeds that for the neutron star stage. As compaction continues, the density increases to the point where, starting at the object's center, the quarks and gluons comprising the nucleonic matter attain asymptotic freedom and deconfine to form a growing sphere of QGP. Further collapse now produces conditions at the center of this sphere such that the preons comprising the quarks themselves deconfine to become a gas of free particles. Transition to a state of negative pressure ("tension") halts the collapse, whereupon the object stabilizes in a quasi-equilibrium state: A hybrid body with a small preon core, a mantle of QGP, and an outer hadronic envelope of nucleonic matter that may be infused with strange-quark nuggets. For simplicity we assume that the transitions between these states of matter are instantaneous once a threshold is reached, so that the boundaries between layers are sharp rather than diffuse.

In this paper, we propose simple, physically plausible stress-energy tensors for every component of this three-layer object, then obtain an exact solution of Einstein's equations within each and match it across that region's boundaries. To this end, for each layer we adopt the simplest possible static, spherically symmetric metric,

$$ds^2 = f.dt^2 - (1/f).dr^2 - r^2 d\Omega^2, \qquad (1)$$

where $f = f(r)$ and $d\Omega^2 \equiv d\theta^2 + \sin^2\theta.d\varphi^2$. (We set $\theta = \pi/2$ and $c = G = 1$ throughout.) This simplifies the Einstein equations and the process of matching metric elements across each boundary, facilitating the construction of a piecewise-defined gravitational potential $f(r)$ that is analytic and continuous from the origin out to infinity.

## 1. The Preon Core

In their 2005 paper, Hansson and Sandin [1] modeled a preon star as a gas of massless fermions (essentially, a perfect fluid with the equation of state $\rho = 3p$) to which they added a bag constant, $B$, to represent the mass-energy contribution resulting from the preons' deconfinement:

$$\rho(r) = 3p(r) + 4B. \tag{2}$$

This will form the basis of our model of the preon core. For the metric (1) with the usual perfect-fluid source tensor and co-moving condition on the 4-velocity,

$$T_{\alpha\beta} = [\rho(r) + p(r)]u_\alpha u_\beta - p(r)g_{\alpha\beta},\ u_t = \sqrt{(g_{tt})},\ u_\alpha u^\alpha = 1, \tag{3}$$

the Einstein equations $G_\alpha^{\ \beta} = -8\pi T_\alpha^{\ \beta}$ reduce to

$$(rf)' - 1 = -8\pi r^2 \rho \tag{4}$$

$$(rf)' - 1 = 8\pi r^2 p \tag{5}$$

$$rf'' + 2f' = 16\pi rp \tag{6}$$

where a prime denotes $d/dr$ and $f$, $\rho$ and $p$ are all functions of $r$. Subtracting equation (4) from (5) yields the Oppenheimer equation, which gives the relationship between pressure and density (the equation of state). An important consequence of the choice of metric (1) is that $G_t^{\ t} = G_r^{\ r}$, so that the left-hand sides of equations (4) and (5) will always be identical; the perfect-fluid Oppenheimer relation then necessarily becomes

$$p(r) = -\rho(r). \tag{7}$$

Requiring that $\rho > 0$ for physically reasonable matter means that the pressure must be negative, making it a tension; it is this that stabilizes the preonic matter against further collapse. Combining relations (2) and (7) now gives the result

$$\rho_1 = B = -p_1, \tag{8}$$

meaning that density and pressure in the core must both be *constant*. (We label all properties of the core with the subscript, 1.) Inserting (8) into either of the Einstein equations (4) or (5) and integrating immediately gives

$$f_1(r) = 1 - (8/3)\pi B r^2, \tag{9}$$

where we have set the constant of integration to zero to eliminate a singularity at the origin. Clearly, the preon core's metric will be well-behaved from the origin to its boundary (at $r = r_1$) as long as

$$r_1 < r_c \equiv [3/(8\pi B)]^{1/2}, \qquad (10)$$

where $r_c$ is the critical radius for which the metric becomes zero. From above, the bag constant obviously determines the core's maximum radius; using the estimate for $B$ from Hansson and Sandin [1], $B \approx 10^4 \text{ TeV}/(\text{fm})^3 = 1.6 \times 10^{42} \text{ J/m}^3$, and rewriting relation (10) in SI units as $r_1 < [3c^4/(8\pi GB)]^{1/2}$, gives $r_1 < 3$ m. The total mass is simply $M = (4/3)\pi r_1^3 \rho_1 = (4/3)\pi r_1^3 (B/c^2)$, again in SI units; the constraint on the maximum radius then limits the core's mass to $M < 5.6 \times 10^{26}$ kg ($\approx 100$ Earth masses). These radius and mass results compare well with previous estimates for a pure preon star [1, 4].

## 2. The QGP Mantle

Assuming that the QGP is in a state of total deconfinement, we may neglect interactions between the quarks and gluons as well as any self-interaction terms. These assumptions are supported by the results of heavy-ion collision experiments carried out at the RHIC only a few years ago, which indicate that the QGP's behavior approaches that of an ideal perfect fluid more closely than any other substance known [5]; we therefore represent the QGP with a perfect-fluid source tensor $(T_{\alpha\beta})_q$ of the same form as (3).

Another assumption we make is that an electric field is present in the QGP (which is itself in a state of tension), perhaps induced by the magnetic flux of the compact object's stellar progenitor during the collapse process [1]. Accordingly, the QGP mantle must contain a spherically symmetric distribution of electrostatic charges whose density, $\sigma(r)$, is expected to decrease with $r$. In our static geometry and co-moving frame, the 4-vector potential for such a distribution of electric charge reduces to

$$A_\alpha = [\varphi(r), 0, 0, 0], \quad E(r) \equiv -d\varphi/dr; \qquad (11)$$

as usual, from $A_\alpha$ is constructed the Maxwell field tensor $F_{\alpha\beta} = A_{\beta,\alpha} - A_{\alpha,\beta}$ (the comma symbolizes ordinary partial differentiation) and from it the familiar energy-momentum tensor

$$(T_{\alpha\beta})_{\text{e-m}} = (1/4\pi)[-F_{\alpha\lambda}F_\beta{}^\lambda + \tfrac{1}{4}g_{\alpha\beta}F_{\lambda\nu}F^{\lambda\nu}]. \qquad (12)$$

The total energy-momentum tensor for the QGP is then

$$T_{\alpha\beta} = [\rho(r) + p(r)]u_\alpha u_\beta - p(r)g_{\alpha\beta} + (1/4\pi)[-F_{\alpha\lambda}F_\beta{}^\lambda + \tfrac{1}{4}g_{\alpha\beta}F_{\lambda\nu}F^{\lambda\nu}]. \qquad (13)$$

Substituting relations (7) and (11) into the electromagnetic component of $T_{\alpha\beta}$, the Einstein equations for the QGP become:

$$(rf)' - 1 = -r^2(8\pi\rho + E^2), \tag{14}$$

$$(rf)' - 1 = r^2(8\pi p - E^2), \tag{15}$$

$$rf'' + 2f' = 2r(8\pi p + E^2). \tag{16}$$

These equations comprise three independent relations for the three unknown functions $f(r)$, $p(r)$ and $E(r)$; recall that $\rho(r) = -p(r)$. The Maxwell equations will introduce a fourth unknown, the electric charge density $\sigma(r)$.

Two sets of conservation equations apply here. The first is $T_\alpha{}^\beta{}_{;\beta} = 0$, where as usual the semicolon symbolizes a covariant derivative. With our chosen metric, the requirement that the covariant divergence of the energy-momentum tensor (13) vanishes reduces to

$$8\pi p' - (1/r^4)[r^4 E^2]' = 0 \tag{17}$$

after applying the Oppenheimer relation (7). This is a form of the well-known Oppenheimer-Volkoff (O-V) equation of hydrostatic equilibrium; and since it can be derived from the Einstein equations, we shall use it in place of one of them—relation (16)—to find the pressure once we have an expression for the electric field. The second set of conservation relations is the Maxwell equations, $F^{\alpha\beta}{}_{;\beta} = 4\pi J^\alpha$, where $J^\alpha \equiv \sigma(r)u^\alpha$ is the electric 4-current density. The only nonzero component of these equations is

$$[r^2(-\varphi)']' = 4\pi r^2 \sigma(r) u^t \tag{18}$$

which, using (11) and $u_t = \sqrt{(g_{tt})} = \sqrt{f(r)}$ and integrating once, gives

$$r^2 E(r) = \int 4\pi r^2 [\sigma(r)/\sqrt{f(r)}] dr. \tag{19}$$

We wish to turn the right-hand side of (19) into a volume integral of charge density, in order that the electric field shall have the expected form $E(r) = q(r)/r^2$ (where $q$ is the electric charge contained within coordinate radius $r$). This is accomplished by means of an ansatz employed by Tiwari et al. [6] which we write as

$$\sigma_2(r) = \sigma_2 . \sqrt{f_2(r)}, \tag{20}$$

where $\sigma_2$ is the maximum value of the charge density (located at the inner boundary of the QGP mantle, $r = r_1$). Carrying out the integration in equation (19) and choosing the constant of integration so that $E_2(r_1) = 0$ then yields

$$E_2(r) \equiv q_2(r)/r^2 = (4/3)\pi\sigma_2(r^3 - r_1^3)/r^2. \tag{21}$$

(We identify all properties of the QGP mantle with the subscript, 2.) Inserting this expression into the O-V equation (17) and integrating gives the pressure—and, from the Oppenheimer relation (7), the density—as

$$p_2(r) = -\rho_2(r) = \chi_2(r) - \chi_2(r_1) + p_1(r_1), \tag{22}$$

where $\chi_2(r) = (2/3)\pi(\sigma_2^2/r)(r^3 + 2r_1^3)$ and we have applied the boundary condition $p_2(r_1) = p_1(r_1) \equiv B$. To complete the solution, the Einstein equation (15) is integrated and the boundary condition $f_2(r_1) = f_1(r_1)$ is used to obtain the metric function inside the mantle:

$$f_2(r) = 1 - (r_1/r)[f_1(r_1) - 1] + (1/r)\Phi_2(r) - (1/r)\Phi_2(r_1), \tag{23}$$

where

$$\Phi_2(r) \equiv \int r^2 \{8\pi p_2(r) - [E_2(r)]^2\} dr. \tag{24}$$

For the metric to be well-behaved throughout the entire QGP mantle, it must not go to zero; writing (23) out explicitly as a polynomial, by using (9) and inserting (21) and (22) into (24) and integrating, turns this requirement into the condition

$$x^2 - \beta^2 x^4 + \alpha^2 r_1^4 (2x^6 - 15x^4 + 20x^3 - 12x + 5) > 0, \tag{25}$$

in which we have defined the dimensionless variable $x = r/r_1$ and two constants (in SI units):

$$\alpha^2 = (16\pi^2/45)(k_C G/c^4)\sigma_2^2, \quad \beta^2 = (r_1/r_c)^2 \tag{26}$$

where $k_C$ is Coulomb's constant. Because $\beta < 1$ since $r_1 < r_c$, it seems reasonable to suppose that $\alpha$ should not differ very much from unity in order that the transition of the metric function across $r_1$ shall be smooth as well as continuous. If we assume that the core radius is close to its maximum value ($\beta = 0.9$, say), then for $\alpha \approx 0.1$ (and smaller) we find from (25) that the metric remains positive for all $r$; but for $\alpha \approx 0.3$ and up to the order of unity, the metric goes to zero at around $x = 1.67$. For this latter scenario, we can estimate an upper limit for the mantle's outer radius $r_2$ and, from (26), the maximum value of its charge density: $r_2 < 4.5$ m, and $\sigma_2 \approx 2 \times 10^{16}$ C/m$^3$. From (21), using $r_1 \approx 3$ m and $r_2 \approx 4.5$ m, the electric field at the outer boundary of the mantle is then $E_2(r_2) \approx 2 \times 10^{27}$ V/m. We note that the upper limit for the electric field of a pure preon star, of mass 100 Earth masses and radius of 1 m, has been estimated to be $10^{34}$ V/m [1].

### 3. The Hadronic Envelope

Assuming that this outer layer is composed of matter similar to the nucleonic fluid inside neutron stars, perhaps intermixed with strange nuggets (as has been proposed for the crusts of strange stars [7]), we may model it as a perfect fluid; its equation of state is still $p(r) = -\rho(r)$, which has also been postulated to hold in the outer layer of a quark star [7]. While it is generally accepted that matter inside known compact objects is essentially charge-neutral, it occurs to us that the tremendous electric field strength (as estimated above) due to the charged QGP at the mantle-envelope boundary could very well produce a charge within the envelope by the classical electrostatic method of induction: Charges of one polarity would be attracted and concentrated towards the inner boundary at $r_2$, while many (or perhaps all) of the opposite charges would be pushed completely out of the envelope and into interstellar space. If the fraction of charges so affected is $\eta$, the relationship between the total charge in the mantle and that induced in the envelope will be

$$Q_3 = -\eta Q_2, \; 0 < \eta \leq 1. \tag{27}$$

(All properties of the envelope carry the subscript, 3.) Defining the electric charge density for the envelope as $\sigma_3(r) = \sigma_3 \cdot \sqrt{f_3(r)}$ where $\sigma_3$ is the charge density at $r = r_2$, then carrying out the integration of (19) exactly as was done for the mantle and choosing the constant of integration so that $E_3(r_2) = E_2(r_2)$, yields

$$E_3(r) = (4/3)\pi[\sigma_3(r^3 - r_2^3) + \sigma_2(r_2^3 - r_1^3)]/r^2. \tag{28}$$

From (21), the total charge in the mantle is $Q_2 = (4/3)\pi\sigma_2(r_2^3 - r_1^3)$; similarly, the total charge in the envelope will be $Q_3 = (4/3)\pi\sigma_3(R^3 - r_2^3)$, where $R$ is the radius of the star. Combining these with (27) now gives the relationship between the maximum charge densities in these regions:

$$\sigma_3/\sigma_2 = -\eta(r_2^3 - r_1^3)/(R^3 - r_2^3). \tag{29}$$

Next, inserting (28) into the O-V equation (17), integrating, and requiring that $p_3(r_2) = p_2(r_2)$ across the mantle-envelope boundary gives

$$p_3(r) = -\rho_3(r) = \chi_3(r) - \chi_3(r_2) + p_2(r_2), \tag{30}$$

where $\chi_3(r) = (2/3)\pi(\sigma_3^2/r)(r^3 + 2r_2^3) - (4/3)\pi(\sigma_2\sigma_3/r)(r_2^3 - r_1^3)$. The Einstein equations for the envelope are the same as those for the mantle, i.e., (14) – (16), but with $E^2$ replaced by $(E_3)^2$. Integrating (15) and applying the boundary condition $f_3(r_2) = f_2(r_2)$ then gives the metric element for the envelope as

$$f_3(r) = 1 - (r_2/r)[f_2(r_2) - 1] + (1/r_2)\Phi_3(r) - (1/r)\Phi_3(r_2), \tag{31}$$

where, except for the subscript change, $\Phi_3(r)$ is identical to (24) and can be written explicitly as a polynomial in $r$ in the same manner.

## 4. Surface Boundary Conditions

At the object's surface, $r = R$, both pressure and density must vanish. From (30) and (22), this requirement becomes

$$\chi_3(R) - \chi_3(r_2) + \chi_2(r_2) - \chi_2(r_1) - B = 0; \tag{32}$$

(29) then allows us to eliminate $\sigma_3$ when writing this condition out in full to give a 6th-order polynomial, from which $R$ may readily be calculated for selected values of $r_1$, $r_2$, $B$, $\sigma_2^2$ and $\eta$. Settting $\eta = 1$, (32) may be written as

$$(y^3 - 1)^2 - (a/b)(2y^3 - 3y^2 + 1) = 0 \tag{33}$$

where $y = R/r_2$, $a = (1/r_2^3)(r_2^3 - r_1^3)^2$ and $b = (r^3 + 2r_1^3 - 3r_1^2 r_2) - 3r_2 B/(2\pi k_C \sigma_2^2)$ in SI units. Straightforward graphical investigation of (33) shows that for a given value of the preon core radius $r_1$, solutions of (32) only exist if $r_2$ is no more than about 1 m larger than $r_1$—that is, the QGP mantle may at most be around 1 m thick. From (10) we have that $r_1 < r_c$, so $r_1 = 0.95 r_c = 2.85$ m is a close approximation for the maximum radius of the preon core. Inserting this into (33) yields a solution only if $r_2$ does not exceed 3.45 m, so that here the QGP mantle is at most 60 cm thick; (33) then gives the star's maximum radius as $R = 3.457$ m and, thus, the hadronic envelope's thickness as a mere 7 mm. If $r_1$ and $r_2$ are now taken to be smaller (1.5 m and 2 m respectively, say), we find that the star's radius decreases (to 2.013 m) and the envelope's thickness grows (to 13 mm). These are general properties of solutions of (33): smaller values of $r_1$ generate smaller values of $R$, suggesting that the star's overall radius is determined by the size of its preon core, while the hadronic envelope's thickness increases as the preon core's radius decreases. (We shall say more about this later.)

A second boundary condition is that the interior metric element $f_3(r)$ must match its exterior equivalent at the star's surface, $r = R$. For the electro-vacuum spacetime surrounding the star, the obvious choice is the Reissner-Nordstrøm metric

$$ds^2 = f_{\text{R-N}}.dt^2 - (1/f_{\text{R-N}}).dr^2 - r^2 d\Omega^2 , \quad f_{\text{R-N}}(r) = 1 - r_S/r + Q^2/r^2, \tag{35}$$

where $r_S$ is the Schwarzschild radius and $Q = Q_2 + Q_3$. Combining (21) and (28) and using (29) to eliminate $\sigma_3$ gives

$$Q = (4/3)\pi(1 - \eta)\sigma_2(r_2^3 - r_1^3)^2; \tag{36}$$

the thinness of the envelope, as found above, should facilitate the expulsion of charges from it due to the Coulomb force from the charged QGP mantle, so we may expect $\eta$ to be close to 1 and, therefore, the star's global electric charge (36) to be close to zero. The condition $f_3(R) = f_{\text{R-N}}(R)$ then gives the Schwarzschild radius as

$$r_S = R[1 - f_3(R)] + Q^2/R^2; \tag{37}$$

requiring that $R > r_S$ so the star does not collapse through its Schwarzschild horizon imposes the condition $f_3(R) > Q^2/R^2$, which guarantees that both $f_3(r)$ and $f_{\text{R-N}}(r)$ will be well-behaved everywhere within their domains of validity. Finally, we use the definition $r_S \equiv 2GM/c^2$ (in SI units) with the (maximum) value of $R$ found from (33) to estimate the star's maximum total gravitational mass as $M_{max} \approx 2 \times 10^{27}$ kg $\approx 300$ Earth masses (or one Jupiter mass); the preon core then constitutes around one-third of the star's total mass.

## 5. Discussion

Two aspects of the material presented so far warrant further comment:

Firstly, it is clear that the model presented here does not represent an object's terminal state, but a stage in the thermal evolution of an ultra-compact object that began either as a post-supernova stellar remnant or as a product of processes in the early Universe. The object's outer regions would cool over time and reconfine to form a QGP; later, the outer layer of the QGP, in turn, would become cool enough to form a layer of hadronic matter. As the object continues to lose energy—however slowly—to its environment, it is logical to expect that its hadronic envelope will grow in thickness while its QGP mantle and preon core will shrink. All of this is indicated by the behavior of solutions to relation (33), which also predicts that the star's overall radius should decrease as it cools. Finally, we expect the object to attain a fully hadronic state (but still possessing negative pressure), or perhaps become a stable, cold QGP with a nucleon crust [8], cooling over time to become some form of black dwarf.

The second point we wish to discuss is the assumption of a state of tension (negative pressure) in each layer of the hybrid star, needed so that gravity is opposed and equilibrium established. This is physically plausible for preonic matter due to its bag energy, but it is not (yet) clear how to apply other negative-pressure mechanisms (such as dark energy and vacuum energy) to other types of matter. This remains perhaps the most speculative feature of the model constructed here.

## 6. Conclusion

We conclude by summarizing what was accomplished in this study: Adopting the simplest possible non-trivial metric architecture facilitated the construction of a three-layer general-relativistic model for a compact object featuring simple, piecewise-continuous, well-behaved polynomial expressions for all relevant fields. Starting with values for just two physical parameters—the preon bag constant $B$ found in the literature, and the QGP's maximum electric charge density $\sigma_2$ deduced here—the basic properties of each layer of the star were calculated and, where applicable, found to compare very well with estimates published in other studies. We were also able to estimate the star's overall radius and total gravitational mass; these, too, agree very closely with values found by other investigators. Finally, we were able to gain some insight into the object's likely evolution over time.

## References

[1] J. Hansson and F. Sandin, Phys. Lett. **B616**, 1 (2005), [astro-ph/0410417].

[2] F. Sandin and J. Hansson, Phys. Rev. D **76**: 125006 (2007), [astro-ph/0701768].

[3] J. Hansson, Acta Phys. Polon. **B38**, 91 (2007), [astro-ph/0603342].

[4] J. E. Horvath, Astrophys. and Space Sci. **307**, 419 (2007), [astro-ph/0702288].

[5] W. A. Zajc, Nucl. Phys. **A805**, 283 (2008), [nucl-ex/0802.3552].

[6] R. N. Tiwari, J. R. Rao and R. R. Kanakamedala, Phys. Rev. D **30**, 489 (1984).

[7] P. Jaikumar, S. Reddy and A. W. Steiner, Phys. Rev. Lett. **96**, 041101 (2006), [nucl-th/0507055].

[8] M. G. B. de Avellar and J. E. Horvath, Int. J. Mod. Phys. **D19**, 1937 (2010), [astro-ph/1006.4629].